\documentclass[preprint, showpacs,preprintnumbers,amsmath,amssymb,nofootinbib]{revtex4}
\usepackage{mathrsfs}
\usepackage{amsmath}
\usepackage[colorlinks, linkcolor=blue, citecolor=red, urlcolor=red]{hyperref}
\usepackage{amsfonts}
\usepackage{latexsym}
\usepackage{amsfonts}
\usepackage{graphicx}
\usepackage{epsf}
\usepackage{dcolumn}
\usepackage{bm}
\newcommand{\bea}{\begin{eqnarray}}
 \newcommand{\eea}{\end{eqnarray}}

 \def\gsim{ \lower .75ex \hbox{$\sim$} \llap{\raise .27ex \hbox{$>$}} }
 \def\lsim{ \lower .75ex \hbox{$\sim$} \llap{\raise .27ex \hbox{$<$}} }
\def\/{\over}

\begin{document}

\title{\bf Oscillating universe in massive gravity}

\author{ Kaituo Zhang$^1$, Puxun Wu$^2$, Hongwei Yu$^{1,2}$ \footnote{Corresponding
author: hwyu@hunnu.edu.cn}}
\address{ $^1$Department of Physics and Key Laboratory of Low Dimensional
Quantum Structures and Quantum Control of Ministry of Education,
Hunan Normal University, Changsha, Hunan 410081, China\\
$^2$ Center for Nonlinear Science and Department of Physics, Ningbo
University,  Ningbo, Zhejiang 315211, China}

\begin{abstract}
Massive gravity is a modified  theory of general relativity.  In this paper, we study, using a method in which the scale factor changes as a particle in a ``potential", all possible cosmic evolutions in a ghost-free massive gravity.    We find that there exists, in certain circumstances, an oscillating universe  or a bouncing one. If the universe starts at the oscillating region, it may undergo a number of oscillations before it quantum mechanically tunnels to the bounce point and then expand forever. But going back to the singularity from the oscillating region is physically not allowed.   So, the big bang singularity can be successfully resolved.  At the same time, we also find that there exists a stable Einstein static state in some cases. However, the universe can not stay at this stable state past-eternally since it is allowed to quantum mechanically tunnel to a big-bang-to-big-crunch region and end with a big crunch. Thus, a stable Einstein static state universe  can not be used to avoid  the big bang singularity in massive gravity.
\end{abstract}


\maketitle
\section{Introduction}
The current accelerated cosmic  expansion was discovered firstly from the Type Ia supernovae data~\cite{Perlmutter1999, Riess1998} and further confirmed by many other observations, including the cosmic microwave background radiation~\cite{Spergel}, the large scale structure~\cite{Eisenstein,Tegmark}, and so on. This discovery broke our common belief  that the universe should be undergoing a decelerated expansion. A possible explanation for it among other such as the cosmological constant and the dynamical dark energy is  that the theory of general relativity is no longer valid on the cosmological scale and needs to be modified.  As a result, many modified gravity theories have been  proposed to explain the   present accelerated cosmic expansion without the need of the mysterious  dark energy. Among them, the Dvali-Gabadadze-Porrati (DGP) model~\cite{DGP} is a very interesting one since it not only admits a self-accelerating solution with only pressureless matter, but also allows the graviton to have a small mass on the cosmological scale.

Actually, about eighty years ago, Fierz and Pauli~\cite{Fierz} first tried to build a theory of massive gravity.
However, the  linear Fierz-Pauli theory can not recover the linearized Einstein gravity in the limit of zero graviton mass and can not pass the solar system tests due to the van Dam-Veltman-Zakharov  (vDVZ) discontinuity~\cite{van}.  With the help of Vainshtein mechanism, the introduction of  nonlinear interactions  can cure this discontinuity~\cite{Vainshtein};  unfortunately, at the same time,  the nonlinear terms also yield the Boulware-Deser (BD) ghost since  more than two time derivatives are contained in them~\cite{Boulware, Arkani}.  In order to construct a consistent theory, nonlinear terms should be tuned to remove order by order the negative energy state in the spectrum~\cite{Boulware}. Recently, a ghost-free nonlinear theory of massive gravity was constructed successfully by de Rham, Gabadadze and Tolley (dRGT)~\cite{Rham}. (See, however, \cite{Deser} for 
the causality issue of the theory.) It has been found that the dRGT theory can accommodate cosmological solutions with self acceleration~\cite{Amico} and the observational constraint on it has been discussed in \cite{Cardone}.   In addition,  the Einstein static state (ES) universe in this massive gravity theory was analyzed in \cite{Parisi} and it was found that there exist stable ES solutions to avoid the big bang singularity problem. Let us note that the Hawking-Moss instanton in massive gravity has been studied in Ref. \cite{Zhangy}.

If one writes  the Friedmann equation  in a form such that the evolution of the cosmic scale factor can be treated as that of a particle in a potential,  then it is  possible to classify all cosmic evolution types  as  has been successfully done  in the Horava-Lifshitz gravity~\cite{Maeda2010}  and  the DGP braneworld scenario~\cite{Zhang2012}.  In this paper, we plan to study all possible cosmic evolutions  in the massive gravity with this method.    The paper is organized as follows. In Sec. II, we give the Friedmann equation of the massive gravity and define all possible cosmic evolution types. In Sec III, we derive the conditions for the ES solution. We then classify all the cosmic types and give the conditions for them in  Sec IV, V and VI, and conclude in Sec VII.

\section{The Friedmann equation}
We consider the theory of massive gravity proposed in \cite{Nieuwenhuizen}. The action has the form
\begin{eqnarray}
S=\frac{1}{8\pi G}\int \bigg(-\frac{1}{2}R+m^2\mathcal{L}\bigg)\sqrt{-g}d^4x+S_m,
\end{eqnarray}
where $G$ is the Newton gravitational constant, $R$ is the Ricci scalar and $\hbar m/c$ is the graviton mass. In the present paper, we let $S_m$ describe the ordinary matter plus a possible cosmological constant generated by vacuum energy.  $\mathcal{L}
$ is the nonlinear higher derivative terms for the massive graviton  and it is defined as
\begin{eqnarray}
\mathcal{L}=\frac{1}{2}(S^2-S^A_B S^B_A)+\frac{c_3}{3!}\epsilon_{MNPQ}\epsilon^{ABCQ}S^M_A S^N_B S^P_C+\frac{c_4}{4!}\epsilon_{MNPQ}\epsilon^{ABCQ}S^M_A S^N_B S^P_C S^Q_D,
\end{eqnarray}
where $S=S^A_A$,  $c_3$ and $c_4$ are two constants, $\epsilon_{MNPQ}$ is the Levi-Civita tensor density and \begin{eqnarray}  S^A_B=\delta^A_B-\gamma^A_B\;.\end{eqnarray}
Here $\gamma^A_B$ is defined by
\bea \gamma^A_B \gamma_A^C=g^{AC}f_{AB}\eea with $f_{AB}$ being a symmetric tensor field.

The Robertson-Walker (RW) metric for a spatially homogeneous and isotropic universe can be written as
\begin{eqnarray}
ds^2=dt^2-a^2(t)\bigg(\frac{dr^2}{1-kr^2}+r^2d^2\Omega\bigg)\;,
\end{eqnarray}
where $a$ is the cosmic scale factor, $t$ is the cosmic time  and $k=0, \pm 1$ is the constant curvature of
three dimensional space. In massive gravity, Chamseddine and Volkov
show that there exist cosmological solutions where the effect of the graviton mass is equivalent to introducing to the Friedmann equation
a matter source that can consist of several different matter types besides a cosmological constant
term~\cite{Chamseddine2011}
\begin{eqnarray}\label{FEq}
H^2+\frac{k}{a^2}=\frac{m^2}{3}\bigg(4c_3+c_4-6+3C\frac{3-3c_3-c_4}{a}+3C^2\frac{c_4+2c_3-1}{a^2}
\nonumber\\
-C^3\frac{c_3+c_4}{a^3}\bigg)+\frac{8\pi G\rho}{3},
\end{eqnarray}
where $H\equiv\frac{\dot{a}}{a}$ is the Hubble parameter,  $C$ is an integration constant and $\rho$ is the energy density of ordinary matter plus vacuum energy.   Here, besides the cosmological term,   three additional terms in the Friedmann  equation which decay as $1/a$, $1/a^2$ and $1/a^3$ can be viewed as quintessence, gas of cosmic strings, and non-relativistic cold matter respectively. Thus, the massive gravity can explain the present accelerated cosmic expansion. At the same time, it may also play an important role in the very early universe (when $a$ is very small). In the present paper, we plan to study the  all the possible cosmic evolutions in massive gravity  and  whether  the big bang singularity can be avoided. Note that, by employing the familiar canonical quantization procedure in massive gravity for an open cosmic background,  Vakili and  Khosravi found that the big bang singularity can be avoided through a bounce~\cite{Vakili}.   For simplicity,  we only consider a spatially flat universe $k=0$ and a positive constant $C$. In addition, since we are interested in the very early universe, the vacuum energy is assumed to be the only cosmic energy component and then  $\rho$ is a positive constant. Defining   $\Lambda\equiv \frac{8\pi G\rho}{3m^2}$ and rescaling  $a/C\rightarrow a$,  we can re-express the Friedmann equation (Eq.~(\ref{FEq})) as
\begin{eqnarray}\label{FEq2}
H^2=\frac{m^2}{3}\bigg(4c_3+c_4-6+3\Lambda+3\frac{3-3c_3-c_4}{a}+3\frac{c_4+2c_3-1}{a^2}-\frac{c_3+c_4}{a^3}\bigg)\;.
\end{eqnarray}

Let us now write the above  Friedmann equation  into the following form
\begin{eqnarray}
\dot{a}^2+V(a)=0,
\end{eqnarray}
where
\begin{eqnarray}\label{potential}
V(a)=-m^2\bigg[\frac{1}{3}(4c_3+c_4-6+3\Lambda)a^2+(3-3c_3-c_4)a+
\nonumber\\
(c_4+2c_3-1)-\frac{1}{3}(c_3+c_4)\frac{1}{a}\bigg].
\end{eqnarray}
Thus the evolution of the scale factor $a$ can be considered as that of a particle
moving in a ``potential"
 $V$. Obviously, this ``potential"  must
satisfy the condition  $V(a)\leq0$, and this gives the possible
ranges of $a$ as the universe evolves. Since  the values of  $\Lambda$,  $c_3$ and  $c_4$ determine the potential $V(a)$,   we can use them to classify all possible cosmic  types.

All types of the universe in the theory of  massive gravity are:

(1) [Bounce]: If $V(a)\leq0$ for $a\in[a_T,\infty)$ and the equality
holds at $a=a_T$, the spacetime initially contracts from an infinite
scale, and it eventually turns around at a finite scale $a_T$, and
then expands forever.

(2) [Oscillation]:  $V(a)\leq0$ for $a\in[a_{min},a_{max}]$ and the
equality occurs at $a=a_{min}$ and $a=a_{max}$. Thus, the spacetime
oscillates between two finite scales.

(3) [$BB\Rightarrow BC$]: $V(a)\leq0$ for $a\in(0,a_T]$ and the equality holds at $a=a_T$. The universe starts from a big bang (BB) and expands. Eventually it turns around at $a=a_T$ and contracts to a big crunch (BC). $a_T$ is the scale factor where the universe turns around from expansion to contraction.

(4) [$BB\Rightarrow \infty$ or $\infty\Rightarrow BC$]: $V(a)<0$ for any positive values of $a$.  The spacetime starts from a big bang and expands forever, or the spacetime always contracts to a big crunch.

\section{Einstein static state solution}

Since we assume that  the universe is  dominated only by vacuum energy, $\rho$ must be a positive constant, and thus $\Lambda=\frac{8\pi G\rho}{3m^2}
>0$.  The potential (Eq.~(\ref{potential}))   can be re-expressed as
\begin{eqnarray}\label{va}
V(a)=-\frac{m^2}{3a}(4c_3+c_4-6+3\Lambda)\bigg(a^3+\frac{3(3-3c_3-c_4)}{(4c_3+c_4-6+3\Lambda)}a^2+
\nonumber\\
\frac{3(c_4+2c_3-1)}{(4c_3+c_4-6+3\Lambda)}a-\frac{(c_3+c_4)}{(4c_3+c_4-6+3\Lambda)}\bigg),
\end{eqnarray}
which means that $V(a)=0$ yields a cubic equation of $a$.
An ES universe appears if there is a solution $a=a_s>0$ which
satisfies $V(a_s)=0$ and $V'(a_s)=0$. At $a_s$, both the speed of the cosmic
expansion and acceleration are equal to zero and thus the universe
can stay at this point in a long time if it is stable. Differentiating $V(a)$ with
respect to $a$, we have
\begin{eqnarray}\label{v'a}
V'(a)=-\frac{m^2}{3a^2}\bigg(2(4c_3+c_4-6+3\Lambda)a^3+3(3-3c_3-c_4)a^2+(c_3+c_4)\bigg)\;.
\end{eqnarray}
Combining $V(a)=0$ and $V'(a)=0$, we find that an ES  solution  requires a relation between $\Lambda$ and other two model  parameters $c_3$, $c_4$ to hold
\begin{eqnarray}\label{Lambdapm}
\Lambda=\Lambda^{\pm}=\frac{\pm2[1+(c_3-1)c_3+c_4]^{\frac{3}{2}}-(c_3-1)[2+c_3(2c_3-1)+3c_4]}{3(c_3+c_4)^2}\;.
\end{eqnarray}
They give two boundaries  for obtaining an oscillating
universe.
Substituting Eq.~(\ref{Lambdapm}) into the equation  $V(a)=0$ or $V'(a)=0$, one can find that the following static state solutions
\begin{eqnarray}\label{as}
a_s=a_s^{\pm}=\frac{2c_3+c_4-1\pm\sqrt{1+(-1+c_3)c_3+c_4}}{3c_3+c_4-3}\;,
\end{eqnarray}
which is a double root of the equation $V(a)=0$ under the condition
$V'(a)=0$, and the third root is
\begin{eqnarray}\label{at}
a_T=a_T^{\pm}=\frac{c_3+c_4}{-1+2c_3+c_4\pm2\sqrt{1+(-1+c_3)c_3+c_4}}\;.
\end{eqnarray}
If $a_T$ is positive, it corresponds to the radius where the universe turns around or
bounces.

 Since the sign of $4c_3+c_4-6+3\Lambda$, which is the coefficient of the $a^3$ term in the potential, plays a crucial role in determining the shape of the potential $V(a)$,
 we will divide our following discussions into three different cases: $4c_3+c_4-6+3\Lambda>0$, $4c_3+c_4-6+3\Lambda<0$ and $4c_3+c_4-6+3\Lambda=0$.

\section{ the case of $4c_3+c_4-6+3\Lambda>0$}
The condition $4c_3+c_4-6+3\Lambda>0$ gives rise to a constraint on $\Lambda$, i.e., $\Lambda>-\frac{1}{3}(4c_3+c_4-6)$. We find that, when $\Lambda$ takes different values,  the number of real roots  for $V(a)=0$ is different. For example, $\Lambda=\Lambda^+$  allows the existence of three roots, where $\Lambda^+$ is defined in Eq.~(\ref{Lambdapm}), but two of them are  double,  which corresponds to  an unstable ES solution. In order to illustrate our results more clearly, we further divide our discussion into  three subcases, i.e.,  $\Lambda^-\leq\Lambda\leq\Lambda^+$,   $\Lambda<\Lambda^-$ and $\Lambda>\Lambda^+$,  respectively.

\subsection{$\Lambda^-\leq\Lambda\leq\Lambda^+$}

In this case,  $V(a)=0$ yields a cubic equation of $a$, which  has three real roots. Assuming $a_1$, $a_2$ and $a_3$ are three real roots of  this cubic equation, respectively,  Eq.~(\ref{va}) can be expressed as
\begin{eqnarray}
V(a)=-\frac{2m^2}{3a}(4c_3+c_4-6+3\Lambda)(a-a_1)(a-a_2)(a-a_3).
\end{eqnarray}
Since  $\Lambda>0$ and  $(4c_3+c_4-6+3\Lambda)>0$,  the  existence of three real roots requires that $\Lambda$ must satisfy the condition  \bea\label{3difroot}  Max\;\bigg \{ 0, -\frac{1}{3}(4c_3+c_4-6)\bigg \} < \Lambda\leq\Lambda^+\;, \eea  if $\Lambda^-\leq Max\;\{0, -\frac{1}{3}(4c_3+c_4-6)\}$, or
 \bea\label{3difroot2} \Lambda^-\leq \Lambda\leq\Lambda^+\;, \eea  if $\Lambda^- > Max\;\{0,  -\frac{1}{3}(4c_3+c_4-6)\}$. Since the cosmic scale factor must be larger than zero ($a>0$), next, we will  only consider the cases of positive roots.

\subsubsection{three positive roots}
 Using $a_{min}$, $a_{max}$ and $a_T$ to represent  three positive roots of $V(a)=0$ and assuming $0<a_{min}\leq a_{max}\leq a_T$, we have
 \begin{eqnarray}\label{va10}
V(a)&=&-\frac{2m^2}{3a}(4c_3+c_4-6+3\Lambda)(a-a_{min})(a-a_{max})(a-a_T)
\nonumber\\
&=&-\frac{2m^2}{3a}(4c_3+c_4-6+3\Lambda)\bigg(a^3-(a_{min}+a_{max}+a_T)a^2+
\nonumber\\ &&(a_{min}a_{max}+a_Ta_{min}+a_{max}a_T)a-a_{min}a_{max}a_T\bigg).
\end{eqnarray}
Three positive roots mean  that $a_{min}+a_{max}+a_T>0$, $a_{min}a_{max}+a_Ta_{min}+a_{max}a_T>0$, and $a_{min}a_{max}a_T>0$.   Using these conditions and  $4c_3+c_4-6+3\Lambda>0$, and comparing Eq.~(\ref{va}) and Eq.~(\ref{va10}), one can obtain three inequalities:
\bea\label{3IE} 3-3c_3-c_4<0\;, \;\; c_4+2c_3-1>0\;,\;\;\; c_3+c_4>0\;.\eea
 When the above inequalities are satisfied,  $\Lambda^-$ is negative or imaginary. Thus, $\Lambda$ is  required to only satisfy  \bea\label{3+root} Max\;\bigg \{-\frac{1}{3}(4c_3+c_4-6), 0\bigg \}< \Lambda\leq\Lambda^+ .\eea

First, we consider the case of  three different positive roots,  which requires $Max\;\{-\frac{1}{3}(4c_3+c_4-6),0\}<\Lambda<\Lambda^+$, and the  $\Lambda=\Lambda^+$ case will be dealt with separately. From this condition and Eq.~(\ref{3IE}), we find that three different positive roots demand
\begin{eqnarray}\label{osci1}
 && c_3\leq\frac{3}{2} \;,\;\;\; c_3+c_4>3-2c_3 \;, \nonumber
\\ or   && c_3>\frac{3}{2} \;,\;\;\;c_3+c_4>0\;.
\end{eqnarray}

\begin{figure}[htbp]
\includegraphics[width=7cm]{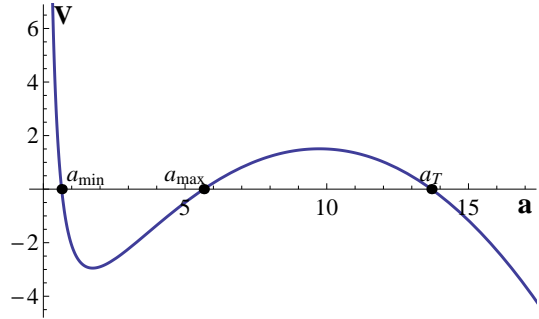}\quad
\caption{\label{Fig1} Potential ${V}(a)$ under the conditions given in Eq.~(\ref{osci1}) and  $Max\;\{-\frac{1}{3}(4c_3+c_4-6),0\}<\Lambda<\Lambda^+$.   An oscillating
universe or a bouncing one is obtained. The constants are
chosen as $m=1$, $c_3=-5$, $c_4=20$, and $\Lambda=2.1$. The radii are $a_{min}=0.64503$,
$a_{max\;}=5.66024$ and $a_T=13.69470$. The period of an oscillation is
$T=5.52532$.}
\end{figure}

As shown in Fig.~(\ref{Fig1}),
$V(a)\leq0$ in $a\in[a_{min},
a_{max}]$ with the equality holding  at
$a=a_{min}$ and $a=a_{max\;}$, and  $a\in[a_T, \infty)$ with $V(a_T)=0$. This means that the
universe oscillates between $a_{min}$ and $a_{max\;}$ or bounces at $a_T$. Thus, if the universe is in the
region $[a_{min}, a_{max\;}]$ initially, it  may undergo
oscillation. After a number of oscillations, it may evolve to the
bounce point $a_T$ through  quantum tunneling but tunneling to the big bang singularity is not allowed. While, if the
universe contracts initially from an infinite scale, it will turn
around at $a_T$ and then expand forever. So, the big bang singularity can be avoided  in this case and a successful emergent universe is achieved.

\begin{figure}[htbp]
\includegraphics[width=7cm]{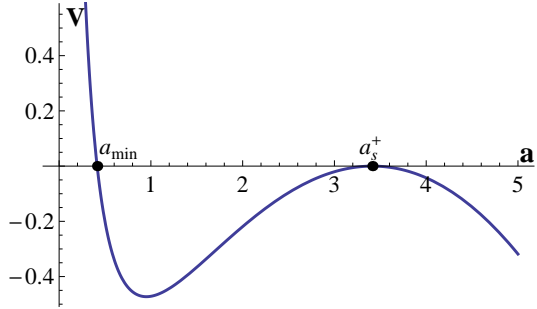}
\caption{\label{Figure3}  Potential $V(a)$ with model parameters satisfying Eq.~(\ref{osci1}) and $\Lambda=\Lambda^+$. There is   an unstable static state universe and a bouncing one. The constants are chosen as $m=1$, $c_3=1$, $c_4=1$, and  $\Lambda=0.47141$.  We obtain  $a_{min}=0.41421$ and  $a_s^+=3.41421$.}
\end{figure}

When $\Lambda=\Lambda^+$ in Eq.~(\ref{3+root}),
 $a_T$ and $a_{max\;}$ coincide with each other, which corresponds to   a double solution. We  represent this double solution  by  $a_s^+$ which is given  in Eq.~(\ref{as}), and find that $c_3$ and $c_4$ are required to satisfy Eq.~(\ref{osci1}), too.  In Fig.~(\ref{Figure3}), we plot the evolutionary curve of $V(a)$ with $\Lambda=\Lambda^+$. It is easy to see that $a_s^+$ is  an unstable ES solution. Therefore, the universe can oscillate
between $a_{min}$ and $a_s^+$, and it can also evolve directly from $a_{min}$
to $\infty$ or evolve to  $\infty$ after some oscillations without
the help of quantum tunneling. If the universe contracts initially from an
infinite scale, it can turn around at $a_s^+$, or pass $a_s^+$ and
 bounce at $a_{min}$, then directly expand forever or do so after a number of oscillations between $a_{min}$ and
$a_s^+$.

\begin{figure}[htbp]
\includegraphics[width=7cm]{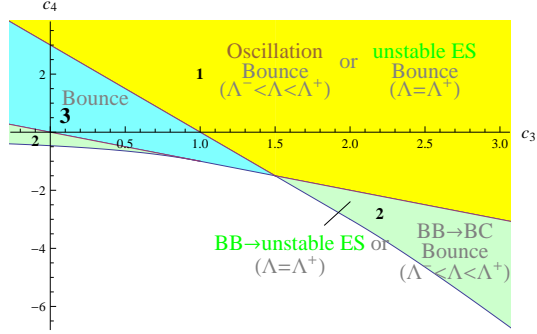}
 \caption{\label{Figo1}   Phase diagram of spacetimes in $(c_3,
c_4)$ plane when $\Lambda^-\leq \Lambda\leq \Lambda^+$. An oscillating universe is
found in Region 1 and a bouncing one is obtained  in Regions 1, 2, and
3.}
\end{figure}

 In Fig.~(\ref{Figo1}), we plot the phase diagram of spacetimes in $(c_3,
c_4)$ plane. Three  positive roots  restrict $c_3$ and $c_4$ to Region 1.

\subsubsection{two positive roots}

In this case, two of three real  roots are positive and one of them is negative. We assume  $a_{1}<0 <a_{2}\leq a_3$, and  set $a_2=a_{T1}$ and $a_3=a_{T2}$ since $a_2$ and $a_3$ are two bouncing points. Then, we have
\bea\label{3product2+1-} a_{1}a_{T1}a_{T2}=\frac{(c_3+c_4)}{(4c_3+c_4-6+3\Lambda)}<0\;. \eea
Of course,   this condition corresponds to two different cases: three negative roots or two positive roots and one negative one. In order to distinguish these two cases,   we further  consider the signs of $a_1+a_{T1}+a_{T2}$ and $a_1a_{T1}+a_{T1}a_{T2}+a_1 a_{T2}$. When $a_1+a_{T1}+a_{T2}\geq0$, there is at least one positive root,  which must correspond to the case of two positive roots and one negative one. When $a_1+a_{T1}+a_{T2}<0$, three negative roots lead to $a_1a_{T1}+a_{T1}a_{T2}+a_1 a_{T2}>0$, while,  $a_1<0<a_{T1}\leq a_{T2}$ implies that $a_1a_{T1}+a_{T1}a_{T2}+a_1 a_{T2}=a_1(a_{T1}+a_{T2})+a_{T1}a_{T2}<-(a_{T1}+a_{T2})^2+a_{T1}a_{T2}=-(a_{T1}+\frac{1}{2}a_{T2})^2-\frac{3}{4}a_{T2}^2<0$.
Thus,  the conditions for  two positive roots and a negative one are:  \begin{eqnarray}\label{2+1-}
&& c_3+c_4<0\;, \;\;-(3-3c_3-c_4)\geq0\;,
\nonumber \\
 or   && c_3+c_4<0\;, \;\;  -(3-3c_3-c_4)<0\;, \;\;c_4+2c_3-1<0\;.
\end{eqnarray}
Same as in the previous subsection, $\Lambda^-$ is negative or imaginary when the above inequalities are satisfied, so, $\Lambda$ is required to only obey
\bea\label{2+root} Max\;\bigg \{-\frac{1}{3}(4c_3+c_4-6),0\bigg \}<\Lambda\leq\Lambda^+\; .\eea

We first consider $Max\;\{-\frac{1}{3}(4c_3+c_4-6),0\}<\Lambda<\Lambda^+$, which corresponds to the case of two different positive roots. Combing this condition with Eq.~(\ref{2+1-}), we find that there exist two different positive roots and one negative one when
\begin{eqnarray}\label{betw2}
&&c_3<1 \;,\;\;3-c_3-2\sqrt{3-3c_3+c_3^2}<c_3+c_4<0\;,
\nonumber \\
&& \frac{3}{2}<c_3\leq2 \;,\;\; 3-2c_3<c_3+c_4<0\;,
\nonumber\\
or && c_3>2\;,\;\;3-c_3-2\sqrt{3-3c_3+c_3^2}<c_3+c_4<0\;.
\end{eqnarray}
Above conditions correspond to Region 2 in  Fig.~(\ref{Figo1}) .

\begin{figure}[htbp]
\includegraphics[width=7cm]{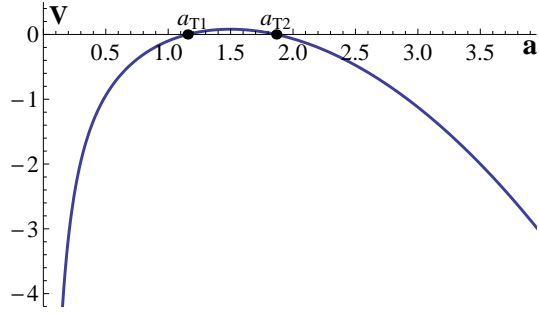}\quad
\caption{\label{F2}  The potential ${V}(a)$ for a $BB\Rightarrow BC$
universe or a bouncing one with $Max\;\{-\frac{1}{3}(4c_3+c_4-6),0\}<\Lambda<\Lambda^+$ and model parameters in Region 2 of Fig.~(\ref{Figo1}).  The constants are
chosen as $m=1$, $c_3=3$, $c_4=-5$, and $\Lambda=0.1$. The radii are $a_{T1}=1.15556$
and $a_{T2}=1.86572$.}
\end{figure}

Fig.~(\ref{F2}) shows the evolution of the potential ${V}(a)$ with
the model parameters in Region 2 of
Fig.~(\ref{Figo1}). From this figure, we find that a $BB\Rightarrow BC$
universe or a bouncing one is
obtained since $V\leq 0$ in $a\in (0, a_{T1}]$ and $[a_{T2},\infty)$ with $V=0$ occurring   at $a=a_{T1}$ and $a=a_{T2}$. Thus, if the universe initiates from a big bang, it can expand to $a_{T1}$. It then turns over at $a_{T1}$ and ends with a big crunch.  In addition, it is also possible that the universe quantum tunnels to $a_{T2}$ directly from $a_{T1}$ and then expands forever. If the universe contracts initially from infinity, a bounce will occur at $a_{T2}$.

\begin{figure}[htbp]
\includegraphics[width=7cm]{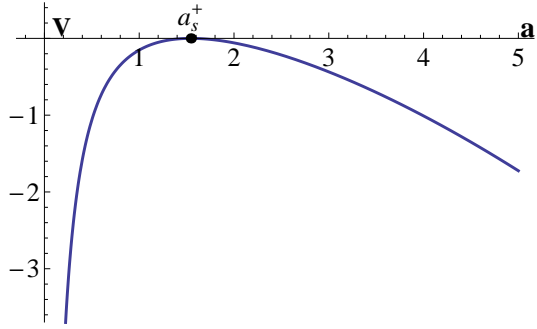}
\caption{\label{Figure5} Potential $V(a)$ with $c_3$ and $c_4$
in Region 2 of Fig.~(\ref{Figo1}), and  $\Lambda=\Lambda^+$.  A big bang
universe and an unstable ES solution are obtained. The constants are chosen as $m=1$, $c_3=3$, $c_4=-6.3$ and $\Lambda=0.15217$. We find $a_s^+=1.54447$.}
\end{figure}

When $\Lambda=\Lambda^+$,
 $a_{T1}$ and $a_{T2}$ coincide with each other  and a double root $a_s^+$ given in Eq.~(\ref{as}) is obtained. $c_3$ and $c_4$ are required to satisfy Eq.~(\ref{betw2}), too.  As shown in Fig~(\ref{Figure5}), the ES solution $a_s^+$ is unstable.  So, if the universe initiates from big bang, it will expand to an unstable ES universe and then turn over or  expand forever. If the universe contracts from infinity initially, it can bounce at $a_s^+$ or end with a big crunch.

\subsubsection{one positive root}
\begin{figure}[htbp]
\includegraphics[width=7cm]{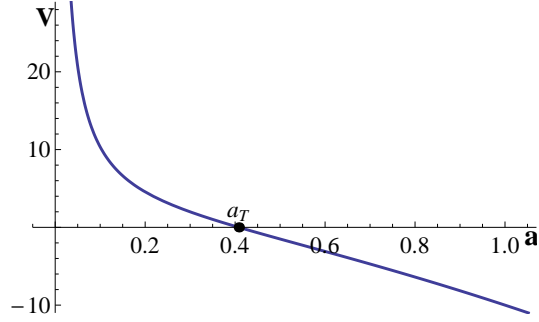}\quad
\caption{\label{Figure6}   Potential ${V}(a)$ with $Max\;\{-\frac{1}{3}(4c_3+c_4-6),0\}<\Lambda<\Lambda^+$ and model parameters in Region 3 of Fig.~(\ref{Figo1}).   A bouncing universe is obtained.  The constants are
chosen as $m=1$, $c_3=-3$, $c_4=6$, and $\Lambda=10$. The turning radius at a bounce is $a_T=0.408248$.}
\end{figure}

For this case, only one of three real roots is positive and two of them are negative.   Assuming that $a_{1},a_{2}<0$ and $a_3=a_T>0$, one has \bea\label{3product1+2-} a_{1}a_{2}a_T=\frac{(c_3+c_4)}{(4c_3+c_4-6+3\Lambda)}>0. \eea
Since  this condition can correspond  to  either three positive roots or  one positive root and two negative ones,  we have to consider  other conditions coming from $a_1+a_2+a_T$ and $a_1a_2+a_2a_T+a_1a_T$. If $a_1+a_2+a_T\leq0$, there is   at least one negative root regardless  of the value of $a_1a_2+a_2a_T+a_1a_T$. Thus,
\bea\label{w1} c_3+c_4 >0\;, \;\; -(3-3c_3-c_4)\leq0\;, \eea
 are sufficient for having one positive root and two negative ones. If $a_1+a_2+a_T>0$, the condition from $a_1a_2+a_2a_T+a_1a_T$ should be added.  When $a_1,a_2,a_T>0$ (three positive roots), $a_1a_2+a_2a_T+a_1a_T>0$. However,  if $a_1<a_2<0<a_T$, then $a_1a_2+a_2a_T+a_1a_T=a_T(a_1+a_2)+a_1a_2<-(a_1+a_2)^2+a_1a_2=-(a_1-\frac{1}{2}a_2)^2-\frac{3}{4}a_2^2<0$. So,
 \bea \label{w2}  c_3+c_4 >0\;, \;\; -(3-3c_3-c_4)>0\;, \;\; c_4 +2c_3-1<0\;, \eea
  give the conditions in obtaining one positive root and two negative ones.  Since $\Lambda^-<0$ if Eq.~(\ref{w1}) or Eq.~(\ref{w2}) is satisfied,  we only  consider
\bea \label{w3} Max\;\bigg \{-\frac{1}{3}(4c_3+c_4-6),0\bigg \}<\Lambda<\Lambda^+\;.  \eea
Here, $\Lambda=\Lambda^+$ is discarded because  in this case there is no double solution. That is, there is no unstable ES solution.
Combining Eqs.~(\ref{w1}) and (\ref{w3}), or Eqs.~(\ref{w2}) and (\ref{w3}), one can find the conditions for the existence of one positive root and two negative ones
\begin{eqnarray}\label{betw3}
c_3<\frac{3}{2}\;\;,\;0<c_3+c_4<3-2c_3\;,
\end{eqnarray}
which is shown as Region 3 in Fig.~(\ref{Figo1}). Fig.~(\ref{Figure6}) shows the evolutionary curve of the potential ${V}(a)$ with  model parameters in  Region 3 in Fig.~(\ref{Figo1}). Apparently,  a bouncing universe is obtained.

\subsubsection{no positive root}
\begin{figure}[htbp]
\includegraphics[width=7cm]{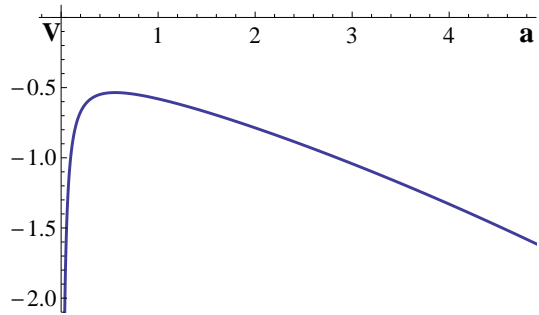}\quad
\caption{\label{Figure7} Potential ${V}(a)$ for the case of no positive root.
A $BB\Rightarrow\infty$ or $\infty\Rightarrow BC$ cosmic type is obtained. The constants are
chosen as $m=1$, $c_3=1.5$, $c_4=-1.7$, and $\Lambda=0.58$.}
\end{figure}

Three real roots are all negative. 
Fig.~(\ref{Figure7}) shows the evolutionary curve of $V(a)$ in this case.  One can see that  the potential is always negative and the cosmic evolution type is $BB\Rightarrow\infty$ or $\infty\Rightarrow BC$.

\subsection{$\Lambda>\Lambda^+$}
\begin{figure}[htbp]
\includegraphics[width=7cm]{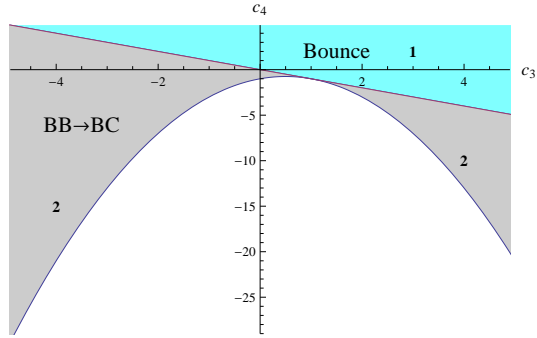}
\caption{\label{Figure8}
 Phase diagram of spacetimes in $(c_3,c_4)$ plane with $\Lambda>\Lambda^+$. A bouncing universe is found in Region 1,  while,  a $BB\Rightarrow BC$ one is obtained in Region 2.}
\end{figure}

In this case, $\Lambda$ must satisfy \bea\label{1realroot+}  \Lambda>Max\;\bigg\{\Lambda^+,-\frac{1}{3}(4 c_3 + c_4 - 6),0\bigg\}\;.\eea
Under this condition, there is only one real root, which can be positive or negative,   and other two roots are a conjugate imaginary pair. In Fig.~(\ref{Figure8}), we show all possible cosmic types in $(c_3, c_4)$ plane.

\subsubsection{one positive root}
\begin{figure}[htbp]
\includegraphics[width=7cm]{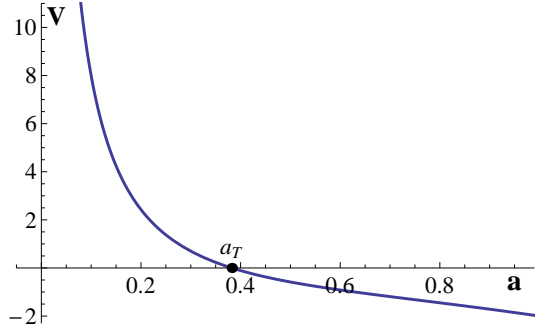}
\caption{\label{Figure9}   Potential $V(a)$ for a bouncing universe with model parameters in Region 1 of Fig.~(\ref{Figure8}) and $\Lambda>\Lambda^+$. The constants are chosen as $m=1$, $c_3=1.5$, $c_4=2$, and $\Lambda=2$. The turning radius at a bounce is $a_T=0.3823023$.}
\end{figure}

Assuming that $a_1$ and $a_2$ are two   conjugate imaginary  roots  and $a_3$ is the only positive one,  and setting $a_3=a_T$,  one has \bea\label{1real+root+} a_1a_2a_T=\frac{(c_3+c_4)}{(4c_3+c_4-6+3\Lambda)}>0.\eea Combining this inequality with   Eq.~(\ref{1realroot+}), we get the condition for one positive root and two  conjugate  imaginary  ones

\begin{eqnarray}\label{p+}
c_3+c_4>0,
\end{eqnarray}
which corresponds to Region $1$ of Fig.~(\ref{Figure8}).
Fig.~(\ref{Figure9}) shows the evolution of the potential ${V}(a)$ with
model parameters in  Region 1 of
Fig.~(\ref{Figure8}). We find that a bouncing universe is
obtained.

\subsubsection{no positive root}
\begin{figure}[htbp]
\includegraphics[width=7cm]{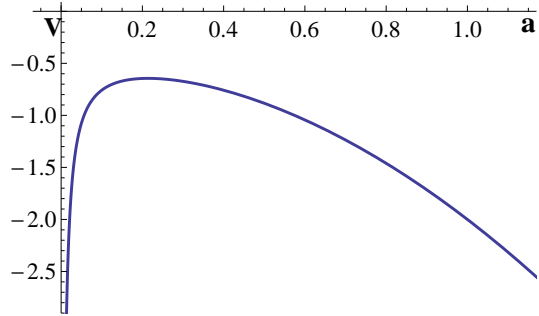}
\caption{\label{Figure10} Potential $V(a)$ for a $BB\Rightarrow\infty$ or $\infty\Rightarrow BC$ universe with $\Lambda>\Lambda^+$ and model parameters in Region 2 of Fig.~(\ref{Figure8}). The constants are chosen as $m=1$, $c_3=1.5$, $c_4=-1.6$, and $\Lambda=2$.}
\end{figure}

In this case, the only real root is   negative, which gives  \bea\label{1real-root+} a_1a_2a_3=\frac{(c_3+c_4)}{(4c_3+c_4-6+3\Lambda)}<0\;. \eea  From the above inequality  and the condition given in Eq.~(\ref{1realroot+}), one has
\begin{eqnarray}\label{p-}
c_3\neq1\;,\;\;-(c_3-1)^2<c_3+c_4<0\;,
\end{eqnarray}
which correspond to Region 2 in Fig.~(\ref{Figure8}).  Fig.~(\ref{Figure10}) shows that the potential is always negative and the cosmic  type is $BB\Rightarrow\infty$ or $\infty\Rightarrow BC$.

\subsection{$\Lambda<\Lambda^-$}
\begin{figure}[htbp]
\includegraphics[width=7cm]{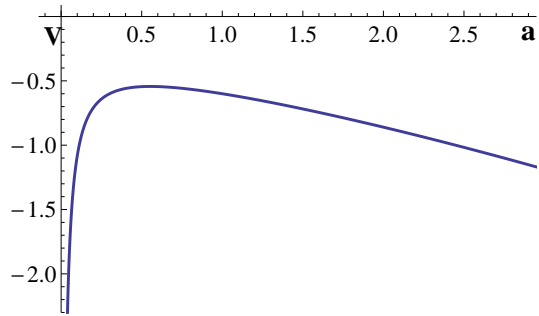}
\caption{\label{Figure11} Potential $V(a)$ under the condition $\{0,-\frac{1}{3} (4 c_3 + c_4 - 6)\}<\Lambda<\Lambda^-$ and model parameters satisfying Eq.~(\ref{m-}).  A $BB\Rightarrow\infty$ or $\infty\Rightarrow BC$ universe  is obtained. The constants are chosen as $m=1$, $c_3=1.5$, $c_4=-1.74$, and $\Lambda=0.6$.}
\end{figure}

It is easy to see that here $\Lambda$ must satisfy
\bea\label{1realroot-} Max\;\bigg\{0,-\frac{1}{3} (4 c_3 + c_4 - 6)\bigg\}<\Lambda<\Lambda^-.\eea
 One can show that there is only one real root and it is negative. Other  two roots are a conjugate imaginary pair. Model parameters $c_3$ and $c_4$ are restricted in the region:
\begin{eqnarray}\label{m-}
1<c_3<2\;,\;\;-(c_3-1)^2\leq c_3+c_4<3-c_3-2\sqrt{3-3c_3+c_3^2}\;.
\end{eqnarray}
The evolution of the potential is shown in Fig.~(\ref{Figure11}). One can see that  the cosmic type is $BB\Rightarrow\infty$ or $\infty\Rightarrow BC$.

 In Tab.~(\ref{Tab1}), we sum up the results obtained in this section.

\begin{table}[!h]
\tabcolsep 3pt \caption{\label{Tab1} Summary of the cosmic type in the $4c_3+c_4-6+3\Lambda>0$ case} \vspace*{-12pt}
\begin{center}
\begin{tabular}{|c|c|c|c|c|c|}
  \hline
  $\Lambda$                                          & $c_3, \; c_4$                                             & Cosmic Type\\ \hline

                                                      &  {$ c_3\leq \frac{3}{2},\; c_3+c_4>3-2c_3 $}      &  Oscillation \\
                                                                  &   {$or~c_3>\frac{3}{2},\; c_3+c_4>0$} & or Bounce \\  \cline{2-3}
  {$Max\;\{-\frac{1}{3}(4c_3+c_4+6), 0\}$}  &   {$c_3<1,\;\;3-c_3-2\sqrt{3-3c_3+c_3^2}<c_3+c_4<0$}  & $BB \Rightarrow BC$ \\
    $<\Lambda<\Lambda^+$   &   {$\frac{3}{2}<c_3\leq2 ,\; 3-2c_3<c_3+c_4<0$} & or Bounce \\
                                            &   {$or~c_3>2,\;3-c_3-2\sqrt{3-3c_3+c_3^2}<c_3+c_4<0$}          &  \\ \cline{2-3}
                     &   {$c_3<\frac{3}{2},\;0<c_3+c_4<3-2c_3$}          &  Bounce \\  \hline
                                                 &  {$ c_3\leq \frac{3}{2},\; c_3+c_4>3-2c_3 $}                  &  Bounce \\
                                              &  {$or~c_3>\frac{3}{2},\; c_3+c_4>0$}                       & Unstable ES \\ \cline{2-3}
        {$\Lambda=\Lambda^+$}                 &   {$c_3<1,\;\;3-c_3-2\sqrt{3-3c_3+c_3^2}<c_3+c_4<0$}          & $BB\Rightarrow\infty$ \\
                                                &   {$\frac{3}{2}<c_3\leq2 ,\; 3-2c_3<c_3+c_4<0$}              &  $BB\Rightarrow BC$ \\
       &   {$or~c_3>2,\;3-c_3-2\sqrt{3-3c_3+c_3^2}<c_3+c_4<0$}  & $\infty\Rightarrow BC$\\
       & & or Bounce \\
       & & Unstable ES  \\ \hline

 $\Lambda>$    &  {$ c_3+c_4>0 $}                  &  Bounce \\  \cline{2-3}
 {$Max\big\{\Lambda^+,-\frac{1}{3}(4 c_3 + c_4 - 6),0\big\}$}   &   {$c_3\neq1,\;\;-(c_3-1)^2<c_3+c_4<0$} & ~ $BB\Rightarrow\infty$ \\
                                            &                                        & or $\infty\Rightarrow BC$ \\  \hline

 {$Max\;\big\{0,-\frac{1}{3} (4 c_3 + c_4 - 6)\big\}$}   &  {$1<c_3<2,$}                 & ~ $BB\Rightarrow\infty$ \\
  $<\Lambda<\Lambda^-$   & {$-(c_3-1)^2\leq c_3+c_4<3-c_3-2\sqrt{3-3c_3+c_3^2}$}     & or $\infty\Rightarrow BC$ \\


  \hline
\end{tabular}
     \end{center}
       \end{table}

\section{The case of $3\Lambda+4c_3+c_4-6<0$}
As in the preceding  section, we divide our discussion  into three different subcases: $\Lambda^-\leq\Lambda\leq\Lambda^+$,   $\Lambda<\Lambda^-$ and $\Lambda>\Lambda^+$, respectively.

\subsection{$\Lambda^-\leq\Lambda\leq\Lambda^+$}
In this case, the equation $V(a)=0$ has three real roots and we  assume them  to be $a_1$, $a_2$ and $a_3$, respectively. Since a positive  $\Lambda$ is considered,   $\Lambda$ must satisfy  \bea\label{b2}0<\Lambda<-\frac{1}{3}(4c_3+c_4-6)\eea and   \bea\label{n3root} \Lambda^-\leq \Lambda\leq \Lambda^+\;.\eea

\subsubsection{three positive roots}

\begin{figure}[htbp]
\includegraphics[width=7cm]{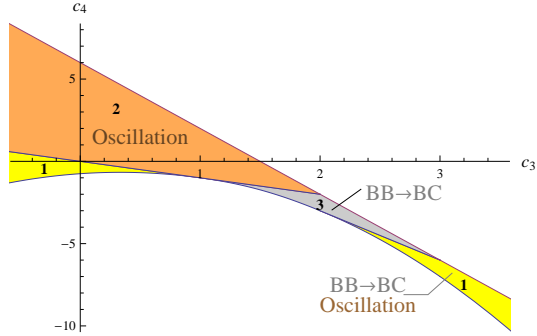}
 \caption{\label{Figure12}   Phase diagram of spacetimes in $(c_3,
c_4)$ plane under the condition of $\Lambda^-<\Lambda<\Lambda^+$. An oscillating universe is
found in Regions 1 and 2, and a $BB\Rightarrow BC$ universe is obtained in Regions 1 and
3.}
\end{figure}

\begin{figure}[htbp]
\includegraphics[width=7cm]{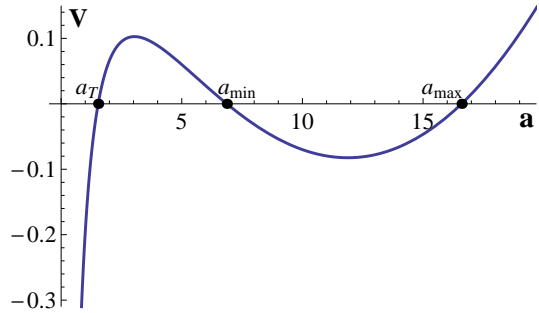}\quad
\caption{\label{Fn1}   Potential ${V}(a)$ for a $BB\Rightarrow BC$
universe or an oscillating one with model parameters in Region 1 of Fig.~(\ref{Figure12}).  The constants are
chosen as $m=1$, $c_3=2.5$, $c_4=-4.6$, and $\Lambda=0.196$. The radii are $a_{T}=1.53548$,
$a_{min}=6.86654$ and $a_{max\;}=16.598$. The period of an oscillation is
$T=107.478$.}
\end{figure}

Letting $a_1=a_{T}$, $a_2=a_{min}$, and $a_3=a_{max}$,  and assuming $0\leq a_T\leq a_{min}\leq a_{max}$,
we can re-express Eq.~(\ref{va})  as
\begin{eqnarray}\label{van1}
V(a)=-\frac{2m^2}{3a}(4c_3+c_4-6+3\Lambda)(a-a_{T})(a-a_{min})(a-a_{max\;})
\nonumber\\=-\frac{2m^2}{3a}(4c_3+c_4-6+3\Lambda)\bigg(a^3-(a_{T}+a_{min}+a_{max\;})a^2+
\nonumber\\(a_Ta_{min}+a_{min}a_{max}+a_T a_{max})a-a_Ta_{min}a_{max\;}\bigg)\;.
\end{eqnarray}
Three positive roots imply that $a_{T}+a_{min}+a_{max\;}>0$, $a_Ta_{min}+a_{min}a_{max}+a_Ta_{max}>0$ and $a_Ta_{min}a_{max\;}>0$. Comparing Eq.~(\ref{va}) and Eq.~(\ref{van1}), one has
\bea\label{n3difrc} 3-3c_3-c_4>0\;, \;\; c_4+2c_3-1<0\;,\;\;c_3+c_4<0.\eea
We first study the case of  $\Lambda\neq\Lambda^+$ and $\Lambda\neq\Lambda^-$, which corresponds to three different positive roots. Combining Eqs.~(\ref{b2}, \ref{n3root}) and Eq.~(\ref{n3difrc}), we obtain that three different positive roots require
\begin{eqnarray}\label{n3osci}
&& c_3<1 \;,\;\; -\frac{3}{4}(c_3-1)^2< c_3+c_4<0\;,
 \nonumber \\
 && 2<c_3\leq3 \;,\;\;-(c_3-1)^2< c_3+c_4<3-2c_3\;,
 \nonumber \\
 or  && c_3>3 \;,\;\;-(c_3-1)^2< c_3+c_4<6-3c_3\;,
\end{eqnarray}
which give Region 1 in Fig.~(\ref{Figure12}). The cosmic type is $BB\Rightarrow BC$ or oscillation, as can be seen from Fig.~(\ref{Fn1}). Thus, if the universe starts from a big bang, it can expand to $a_{T}$, then turn over at $a_{T}$ and end with a big crunch.  If the universe is in the region $a\in[a_{min},a_{max\;}]$ initially, it  may undergo an
oscillation. After some oscillations, it may quantum mechanically tunnel to $a_T$ and end with a big crunch singularity. Therefore,  the classical  singularity still exists.

\begin{figure}[htbp]
\includegraphics[width=7cm]{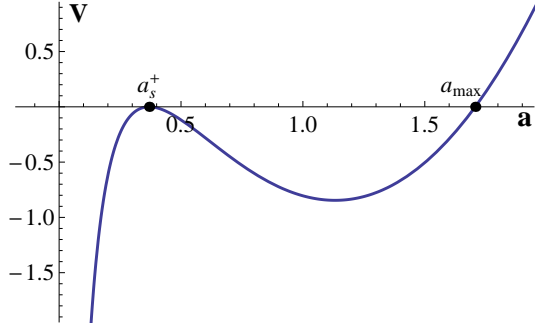}
\caption{\label{Fn4}  The potential $V(a)$ under the condition $\Lambda=\Lambda^+$. An unstable ES solution is obtained. The constants are chosen as $m=1$, $c_3=-1$, $c_4=-1$, and $\Lambda=0.804738$.  The radii are $a_{s}^+=0.369398$ and $a_{max\;}=1.70711$.}
\end{figure}

When $\Lambda=\Lambda^+$ and $\Lambda$ satisfies  $0<\Lambda<-\frac{1}{3}(4c_3+c_4-6)$, $a_{T}$ and $a_{min}$ coincide with each other and forms a double positive root $a_s^+$  defined in Eq.~(\ref{as}). Fig.~(\ref{Fn4}) shows that  $a_s^+$ is an unstable ES solution.  The cosmic type is $BB\Rightarrow BC$. But, the universe can  turn over at $a_s^+$, or $a_{max}$.   From Eq.~(\ref{van1}) and  $0<\Lambda=\Lambda^+<-\frac{1}{3}(4c_3+c_4-6)$, we obtain the conditions on $c_3$ and $c_4$ for an unstable Einstein static state solution:
\begin{eqnarray}\label{nuns1}
&&c_3<1\;,\;\; -\frac{3}{4}(c_3-1)^2<c_3+c_4<3-c_3-2\sqrt{3-3c_3+c_3^2}\;,
\nonumber\\
or && c_3>2\;,\;\;-(c_3-1)^2\leq c_3+c_4<3-c_3-2\sqrt{3-3c_3+c_3^2}\;.
\end{eqnarray}

\begin{figure}[htbp]
\includegraphics[width=7cm]{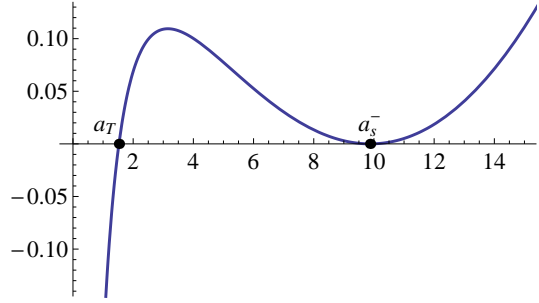}
\caption{\label{Fn5}    Potential $V(a)$  under the condition $\Lambda=\Lambda^-$. A $BB\Rightarrow BC$ universe or a stable  ES one is obtained. The constants are chosen as $m=1$, $c_3=2.5$, $c_4=-4.6$, and $\Lambda=0.195299$. The radii are $a_T=1.52772$ and $a_s^-=9.87298$.}
\end{figure}

If $0<\Lambda=\Lambda^-<-\frac{1}{3}(4c_3+c_4-6)$, there is also a double solution $a_s^-$, which is the coincidence of $a_{min}$ and $a_{max\;}$,   and, as shown in Fig.~(\ref{Fn5}), it is a stable ES solution.   This stable ES solution requires that $c_3$ and $c_4$ satisfy
\begin{eqnarray}\label{nstab1}
&& 2<c_3\leq3\;,\;\;\;\;-(c_3-1)^2\leq c_3+c_4<3-2c_3\;,
\nonumber\\
or && c_3>3\;,\;\;\;-(c_3-1)^2\leq c_3+c_4<-\frac{3}{4}(c_3-1)^2\;.
\end{eqnarray}
Since $V(a)\leq0$ in $a\in(0,a_{T}]$ and $a=a_s^-$,  the cosmic type is $BB\Rightarrow BC$ if the scale factor is less than $a_T$ initially,  or  the universe stays at $a_s^-$.  However,  the universe can not stay at this stable ES past-eternally since  quantum tunneling may drive it into the region  $a\in(0,a_{T}]$. Thus, the big bang singularity can not be avoided although there is a stable ES solution.   As a result,  in  massive gravity the existence of a stable ES solution can not successfully resolve the big bang singularity problem.

\begin{figure}[htbp]
\includegraphics[width=7cm]{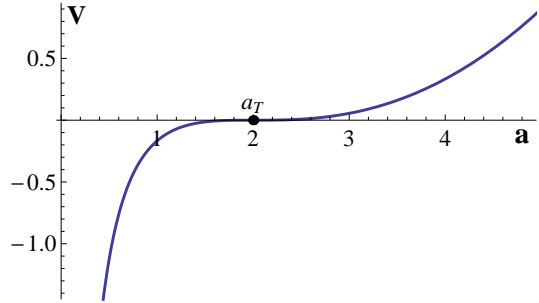}
\caption{\label{Fn6}   Potential $V(a)$ under the condition $\Lambda=\Lambda^+=\Lambda^-$. A $BB\Rightarrow BC$ universe is obtained.
 The constants are chosen as $m=1$, $c_3=3$, $c_4=-7$, and $\Lambda=0.166667$. The radius is $a_T=2$.}
\end{figure}

In addition, there is also a possibility such that $\Lambda=\Lambda^+=\Lambda^-$,  which means that $a_T$, $a_{min}$ and $a_{max}$ merges to form a triple root $a_T$.  Using  \bea\label{n3cubicroot} 0<\Lambda=\Lambda^+=\Lambda^-<-\frac{1}{3}(4c_3+c_4-6)\;,\eea
and Eq.~(\ref{n3difrc}), one can find that the conditions  for a triple root are $c_3>2$, $c_4=-1+c_3-c_3^2$ and $\Lambda=\frac{1}{3(c_3-1)}$.
From Fig.~(\ref{Fn6}),  we can see that the potential $V(a)\leq0$ when $a\in(0,a_T]$, which means that  the cosmic type is $BB\Rightarrow BC$.

\subsubsection{two positive roots}

\begin{figure}[htbp]
\includegraphics[width=7cm]{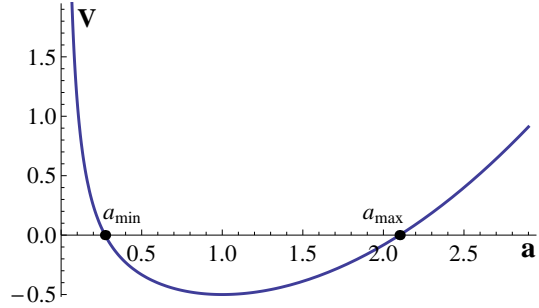}\quad
\caption{\label{Fn2}   Potential ${V}(a)$ for an oscillating
universe with model parameters in Region 2 of Fig.~(\ref{Figure12}).  The constants are
chosen as $m=1$, $c_3=1$, $c_4=-0.5$, and $\Lambda=0.5$. The radii are $a_{min}=0.27255$
and $a_{max}=2.10074$.}
\end{figure}

We assume $a_{1}<0$, and $a_{2},a_3>0$,  and set $a_2=a_{min}$ and $a_2=a_{max}$, so that
 \bea\label{n3product2+1-} a_{1}a_{min}a_{max}=\frac{(c_3+c_4)}{(4c_3+c_4-6+3\Lambda)}<0. \eea
 An analysis similar to that in the previous section leads
  to the conditions for  two positive roots and one negative root as follows:
\begin{eqnarray}\label{n2osci}
c_3<2 \;,\;\;0<c_3+c_4<6-3c_3,
\end{eqnarray}
which determine Region 2 of
Fig.~(\ref{Figure12}). We find that $\Lambda=\Lambda^\pm$ is forbidden since $\Lambda^+>-\frac{1}{3}(4c_3+c_4-6)$ and $\Lambda^-<0$ when $c_3$ and $c_4$ satisfy Eq.~(\ref{n2osci}). This implies that $\Lambda$ is only required to satisfy Eq.~(\ref{b2}).
Fig.~(\ref{Fn2}) shows the evolution of the potential ${V}(a)$ with
the model parameters in Region 2 of
Fig.~(\ref{Figure12}). From Fig.~(\ref{Fn2}), we find that an oscillating universe is
achieved since $V\leq 0$ in $a\in [a_{min}, a_{max}]$ with $V=0$ occurring  at $a=a_{min}$ and $a_{max}$.

\subsubsection{one positive root}
\begin{figure}[htbp]
\includegraphics[width=7cm]{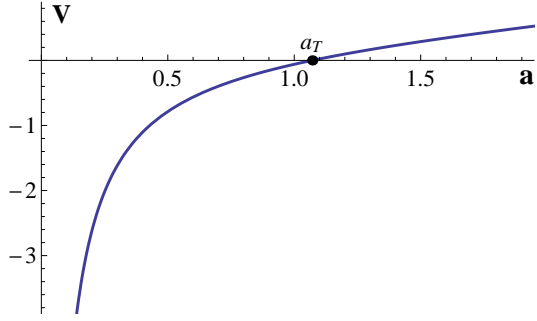}\quad
\caption{\label{Fn3}    Potential ${V}(a)$ for  a $BB \Rightarrow BC$ universe with model parameter in Region 3 of Fig.~(\ref{Figure12}).  The constants are chosen as $m=1$, $c_3=2.5$, $c_4=-4.25$, and $\Lambda=0.06$. and the turning radius is $a_T=1.07118$.}
\end{figure}

Assuming that $a_{1},a_{2}<0$ and $a_3=a_T>0$ and using the  analysis similar to that in the previous section,
 we get that the conditions for one positive root
\begin{eqnarray}\label{n1osci}
&& 1<c_3<2\;,\;\;\;\;3-c_3-2\sqrt{3-3c_3+c_3^2}<c_3+c_4<0\;,
\nonumber \\
or && 2\leq c_3<3\;,\; \;\;\;3-2c_3<c_3+c_4<6-3c_3\;.
\end{eqnarray}
Region 3 of
Fig.~(\ref{Figure12}) shows the allowed values of $c_3$ and $c_4$ for only one positive root.  Fig.~(\ref{Fn3}) shows the evolution of the potential ${V}(a)$. It is easy to see that  a $BB\rightarrow BC$ universe is
realized.

\subsubsection{no positive root}
Since $a_{1},a_{2},a_3<0$, and $4c_3+c_4-6+3\Lambda<0$, one has three inequalities:
\bea\label{n3-root} 3-3c_3-c_4<0, \;\; c_4+2c_3-1<0,\;\;\; c_3+c_4>0.\eea
When $c_3$ and $c_4$ satisfy the above  inequalities,  $\Lambda^-<0$ and $\Lambda^+<0$. Therefore, there is no allowed positive value for $\Lambda$.  This means that this  is not a physically meaningful case.

\subsection{$\Lambda>\Lambda^+$}
This case corresponds to  only one real root $a_3$. Other two roots $a_1$  and $a_2$ are  a conjugate imaginary pair.
 Now $\Lambda$ must satisfy
\bea\label{n1realroot+} Max\;\{0, \Lambda^+\}<\Lambda<-\frac{1}{3} (4 c_3 + c_4 - 6)\eea If this real root $a_3$  is  negative,  then \bea\label{n1real+root-} a_1a_2a_3=\frac{(c_3+c_4)}{(4c_3+c_4-6+3\Lambda)}<0.\eea Considering the condition of $\Lambda$ (Eq.~(\ref{n1realroot+})), we find that there is no solution for Eq.~(\ref{n1real+root-}). Thus, $a_3$ must be a positive  one  and we set $a_3=a_T$, which means
\bea\label{n1real+root+} a_1a_2a_T=\frac{(c_3+c_4)}{(4c_3+c_4-6+3\Lambda)}>0.\eea

\begin{figure}[htbp]
\includegraphics[width=7cm]{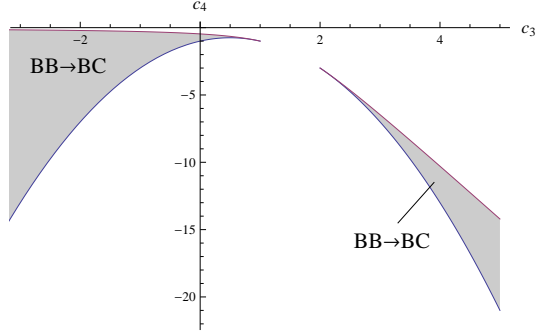}
\caption{\label{Fnd1}
 Phase diagram of spacetimes in $(c_3,c_4)$ plane when $\Lambda>\Lambda^+$. A $BB\Rightarrow BC$ universe is found in the gray region.}
\end{figure}

\begin{figure}[htbp]
\includegraphics[width=7cm]{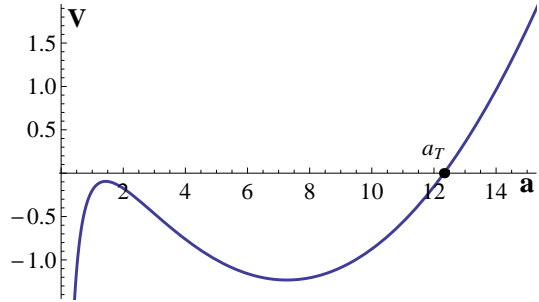}
\caption{\label{Fn7}   Potential $V(a)$ with model parameters in the gray regions of Fig.~(\ref{Fnd1}) and $\Lambda>\Lambda^+$. We obtain a $BB\Rightarrow BC$ universe. The constants are chosen as $m=1$, $c_3=3$, $c_4=-6.75$, and $\Lambda=0.2$. The radius is $a_T=12.3248$.}
\end{figure}

Combining Eq.~(\ref{n1realroot+}) and Eq.~(\ref{n1real+root+}), we obtain that   $c_3$ and $c_4$ must obey
\begin{eqnarray}\label{p+}
    && c_3<1\;,\;\;\;-(c_3-1)^2\leq c_3+c_4<3-c_3-2\sqrt{3-3c_3+c_3^2}\;,\nonumber \\
or && c_3>2 \;,\;\;\;-(c_3-1)^2\leq c_3+c_4<3-c_3-2\sqrt{3-3c_3+c_3^2}\;,
\end{eqnarray}
in order to get one positive root and two imaginary roots which are conjugate to each other.
In Fig.~(\ref{Fnd1}), we   show the allowed values of $c_3$ and $c_4$ for a positive root.   Fig.~(\ref{Fn7}) shows the evolution of the potential ${V}(a)$ with
the model parameters in the gray regions of
Fig.~(\ref{Fnd1}), and a $BB\Rightarrow BC$ universe is
obtained.

\subsection{$\Lambda<\Lambda^-$}
\begin{figure}[htbp]
\includegraphics[width=7cm]{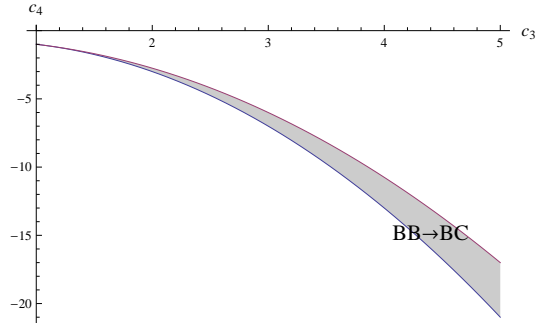}
 \caption{\label{Fnx1} Phase diagram of spacetimes in $(c_3,
c_4)$ plane under the condition $\Lambda<\Lambda^-$. A $BB\Rightarrow BC$ universe is found in the gray region.}
\end{figure}

\begin{figure}[htbp]
\includegraphics[width=7cm]{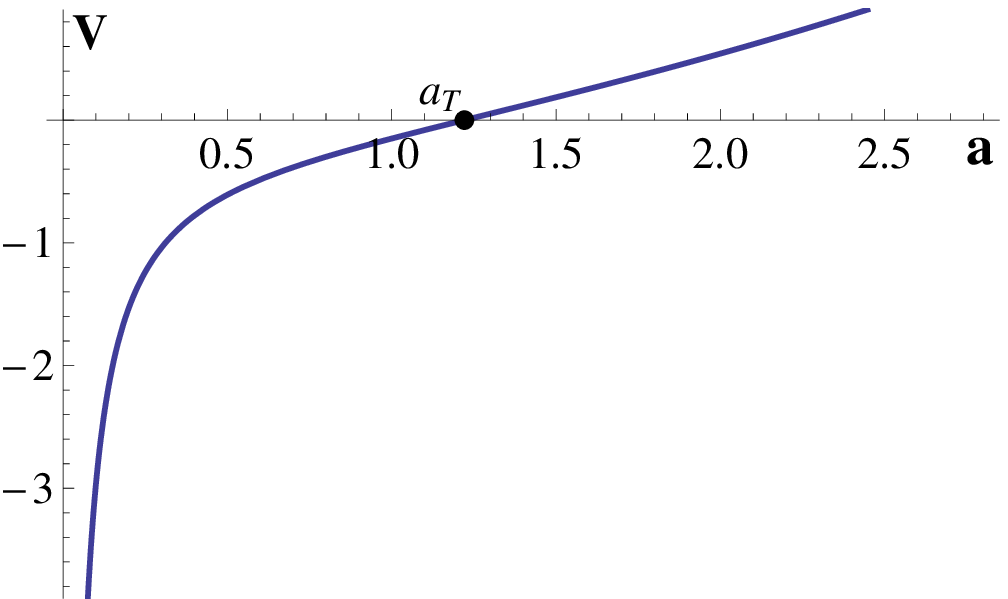}
\caption{\label{Fn8}  Potential $V(a)$ with model parameters in the gray region of Fig.~(\ref{Fnx1}) and $\Lambda<\Lambda^-$. We obtain a $BB\Rightarrow BC$ universe. The constants are chosen as $m=1$, $c_3=2$, $c_4=-2.85$, and $\Lambda=0.15$. The radius is $a_T=1.22118$}
\end{figure}

This case requires
\bea\label{n1realroot-} 0<\Lambda<Min\bigg \{\Lambda^-,-\frac{1}{3} (4 c_3 + c_4 - 6)\bigg\}\;.\eea
Under this condition Eq.~(\ref{n1real+root-}) can't be satisfied, but Eq.~(\ref{n1real+root+}) can. Thus,  there is  also one positive root  $a_3=a_T$ and two conjugate imaginary ones $a_1$ and $a_2$. Combining the condition on $\Lambda$ given in Eq.~(\ref{n1realroot-}) and Eq.~(\ref{n1real+root+}), we get
\begin{eqnarray}\label{nm-}
c_3>1\;,\;\; -(c_3-1)^2\leq c_3+c_4<-\frac{3}{4}(c_3-1)^2\;.
\end{eqnarray}
The allowed region of $c_3$ and $c_4$ is shown in Fig.~(\ref{Fnx1}) and the evolutionary curve of the potential is given in Fig.~(\ref{Fn8}).  We find that  the potential $V(a)\leq 0$ when $0<a\leq a_T$  and the cosmic type is $BB\Rightarrow BC$.

Tab.~(\ref{Tab2}) sums up the results of this section.

\begin{table}[!h]
\tabcolsep 3pt \caption{\label{Tab2} Summary of the cosmic type in the $4c_3+c_4-6+3\Lambda<0$ case} \vspace*{-12pt}
\begin{center}
\begin{tabular}{|c|c|c|c|c|c|}
  \hline
  $\Lambda$                                          & $c_3, \; c_4$                                             & Cosmic Type\\ \hline

                                                      &  {$ c_3<1,\;\; -\frac{3}{4}(c_3-1)^2< c_3+c_4<0 $}      &  $BB\Rightarrow BC$  \\
                                                    &   {$2<c_3\leq3,\;\;-(c_3-1)^2< c_3+c_4<3-2c_3$ }             & or Oscillation \\
           {$Max\;\{\Lambda^-, 0\}<\Lambda<$}            &  {$or~ c_3>3,\;\;-(c_3-1)^2< c_3+c_4<6-3c_3 $}      &  \\ \cline{2-3}
       {$Min\{ -\frac{1}{3}(4c_3+c_4+6),\Lambda^+\}$}           &   {$c_3<2,\;\;0<c_3+c_4<6-3c_3$}  &  Oscillation\\  \cline{2-3}
                                                 &   {$1<c_3<2,\;3-c_3-2\sqrt{3-3c_3+c_3^2}<c_3+c_4<0$}  &  $BB\Rightarrow BC$  \\
                                           &   {$ or~2\leq c_3<3,\; \;\;\;3-2c_3<c_3+c_4<6-3c_3$}          &  \\ \hline

         &{$ c_3<1,\;-\frac{3}{4}(c_3-1)^2<c_3+c_4$}      &  \\
   {$0<\Lambda=\Lambda^+<$}   &{$<3-c_3-2\sqrt{3-3c_3+c_3^2}  $}               & $BB\Rightarrow BC$  \\
              $-\frac{1}{3}(4c_3+c_4+6)$           &  {$or~c_3>2,-(c_3-1)^2\leq c_3+c_4$}  &    Unstable ES        \\
                                                        &      {$<3-c_3-2\sqrt{3-3c_3+c_3^2}$}        &             \\ \hline

  {$0<\Lambda=\Lambda^-<$} &{$ 2<c_3\leq3,\;\;-(c_3-1)^2\leq c_3+c_4<3-2c_3 $}  &  $BB\Rightarrow BC$ \\
  $ -\frac{1}{3}(4c_3+c_4-6)$     &  {$or~c_3>3,\;-(c_3-1)^2\leq c_3+c_4<-\frac{3}{4}(c_3-1)^2$}              &  Stable ES \\ \hline

{$\Lambda=\Lambda^-=\Lambda^+=\frac{1}{3(c_3-1)}$} &{$ c_3>2,\;\;c_3+c_4=-(c_3-1)^2 $}  &  $BB\Rightarrow BC$ \\  \hline

                                                         &{$c_3<1,\;\;-(c_3-1)^2\leq c_3+c_4$}&  \\
{$Max\;\{0, \Lambda^+\}<\Lambda<$}  &{$<3-c_3-2\sqrt{3-3c_3+c_3^2}$}& $BB\Rightarrow BC$ \\
  $-\frac{1}{3}(4c_3 + c_4 -6)$    & {$or~c_3>2 ,\;\;-(c_3-1)^2\leq c_3+c_4$}   &  \\
                                                 & {$<3-c_3-2\sqrt{3-3c_3+c_3^2}$}   &  \\  \hline

 {$0<\Lambda<$}  &  {$c_3>1,\;\; -(c_3-1)^2\leq c_3+c_4<-\frac{3}{4}(c_3-1)^2$} &  $BB\Rightarrow BC$ \\
$Min\big \{\Lambda^-,-\frac{1}{3} (4 c_3 + c_4 - 6)\big\}$ & &\\

  \hline
\end{tabular}
     \end{center}
       \end{table}

\section{The case of $\Lambda=-\frac{1}{3}(4c_3+c_4- 6)$}
In this case Eq.~(\ref{potential}) becomes
\begin{eqnarray}\label{potential2}
V(a)=\frac{m^2}{a}\bigg[-(3-3c_3-c_4)a^2-
(c_4+2c_3-1)a+\frac{1}{3}(c_3+c_4)\bigg]\;.
\end{eqnarray}
Apparently, $V(a)=0$ has two roots:
\begin{eqnarray}\label{cubic0double+}
a_{1}=\frac{3-6c_3-3c_4-\sqrt{3(3+6c_4-4c_3c_4-c_4^2)}}{-6(3-3c_3-c_4)}
\nonumber\\
a_{2}=\frac{3-6c_3-3c_4+\sqrt{3(3+6c_4-4c_3c_4-c_4^2)}}{-6(3-3c_3-c_4)}
\end{eqnarray}
Now  we divide our discussion  into two cases, i.e., $-(3-3c_3-c_4)>0$ and $-(3-3c_3-c_4)<0$.

\subsection{$-(3-3c_3-c_4)>0$}
\begin{figure}[htbp]
\includegraphics[width=7cm]{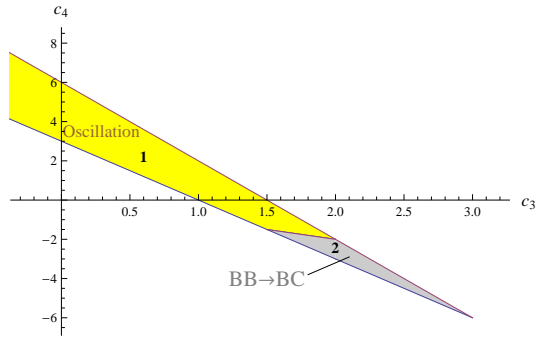}
 \caption{\label{Figurecubic0+}   Phase diagram of spacetimes in $(c_3,
c_4)$ plane under the condition $\Lambda=-\frac{1}{3}(4c_3+c_4- 6)$ and $-(3-3c_3-c_4)>0$. An oscillating universe is
found in Region 1,  and a $BB\Rightarrow BC$ one is obtained in Region 2.}
\end{figure}

\begin{figure}[htbp]
\includegraphics[width=7cm]{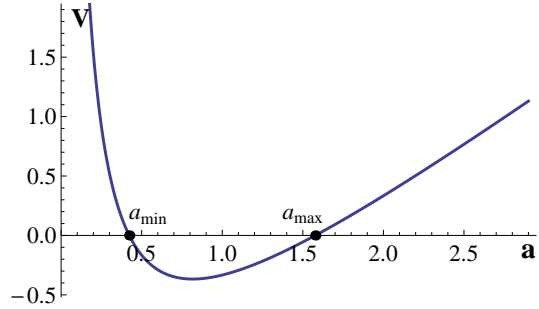}\quad
\caption{\label{Figurez1}    Potential ${V}(a)$ with model parameters in Region 1 of Fig.~(\ref{Figurecubic0+}). An oscillating
universe  is obtained.  The constants are
chosen as $m=1$, $c_3=1$, $c_4=1$, and $\Lambda=\frac{1}{3}$. The radii are $a_{min}=0.42265$
and $a_{max}=1.57735$.}
\end{figure}

If two roots are all positive,  we set $a_2=a_{min}$ and $a_3=a_{max}$,  and find that  $c_3$ and $c_4$ must satisfy $\Delta=(3+c_4(6-4c_3)-c_4^2)>0$, $ c_3 + c_4>0$, $-(3-3c_3-c_4)>0$, $\Lambda=-\frac{1}{3}(4c_3+c_4- 6)>0$, and $-(c_4 + 2 c_3 - 1) < 0$. These lead to
\begin{eqnarray}\label{cubic02+1}
&& c_3\leq \frac{3}{2} \;,\;\;3-2c_3<c_3+c_4<6-3c_3\;,
\nonumber\\
or && \frac{3}{2}<c_3<2 \;,\;\;0<c_3+c_4<6-3c_3\;,
\end{eqnarray}
which is represented as Region 1 in Fig.~(\ref{Figurecubic0+}) where the phase diagram of spacetimes in $(c_3, c_4)$ plane is shown.  Fig.~(\ref{Figurez1}) displays the evolution of the potential with model parameters in Region 1 of Fig.~(\ref{Figurecubic0+}). We can see that
$V(a)\leq0$ in $a\in[a_{min},
a_{max}]$  with the equality holding  at
$a=a_{min}$ and $a=a_{max}$. Thus,  an oscillating universe is obtained.

If $a_{1}=a_{2}$, then $c_4=3-2c_3\pm2\sqrt{3+c_3(c_3-3)}$, which is outside Region 1 of Fig.~(\ref{Figurecubic0+}). Thus,  a stable static Einstein universe can't be obtained in this case.

\begin{figure}[htbp]
\includegraphics[width=7cm]{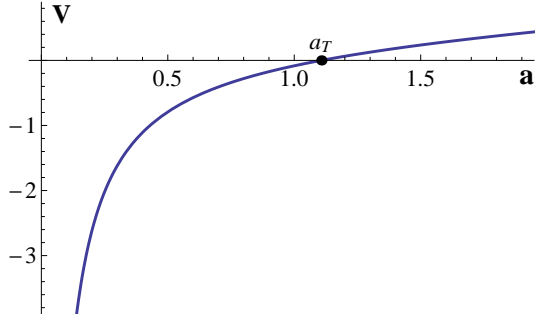}\quad
\caption{\label{Figurez2}   Potential ${V}(a)$ with model parameters in  Region 2 of Fig.~(\ref{Figurecubic0+}). A $BB\Rightarrow BC$ universe is obtained.  The constants are
chosen as $m=1$, $c_3=2.5$, $c_4=-4.25$, and $\Lambda=0.083333$. and the turning radius is $a_T=1.10728$.}
\end{figure}

Now we consider the case of $a_{1}<0$ and $a_{2}=a_T>0$.  We find that  there is a $BB\Rightarrow BC$ cosmic evolution type as  shown in Fig.~(\ref{Figurez2}),  and $c_3$ and $c_4$  satisfy
\begin{eqnarray}\label{cubic02+1}
&& \frac{3}{2}<c_3\leq 2  \;,\;\;3-2c_3<c_3+c_4<0\;,
\nonumber\\
or && 2<c_3<3 \;,\;\;3-2c_3<c_3+c_4<6-3c_3\;,
\end{eqnarray}
 which correspond to Region 2 in Fig.~(\ref{Figurecubic0+}).

If both $a_{1}$ and $a_{2}$ are negative, then $V(a)>0$  in $a\in(0,
\infty)$. So this is not a case of physical significance.

\subsection{$-(3-3c_3-c_4)<0$}
\begin{figure}[htbp]
\includegraphics[width=7cm]{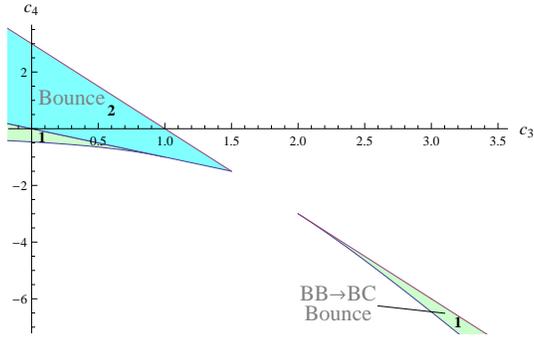}
 \caption{\label{Figurecubic0-}   Phase diagram of spacetimes in $(c_3,
c_4)$ plane under the conditions  $\Lambda=-\frac{1}{3}(4c_3+c_4- 6)$ and $-(3-3c_3-c_4)<0$. A bouncing universe is
found in Regions 1 and 2, and  a $BB\Rightarrow BC$ universe is obtained in Region 1.}
\end{figure}

\begin{figure}[htbp]
\includegraphics[width=7cm]{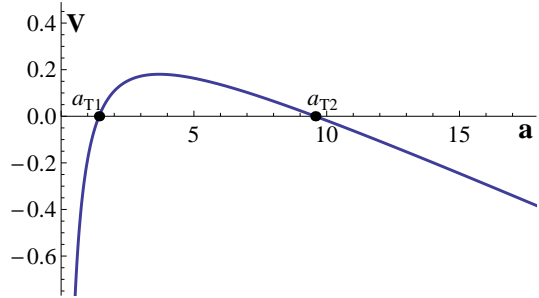}\quad
\caption{\label{Figurez3}  Potential ${V}(a)$ with model parameters in  Region 1 of Fig.~(\ref{Figurecubic0-}). A $BB\Rightarrow BC$ universe or a bouncing one is found.  The constants are chosen as $m=1$, $c_3=2.5$, $c_4=-4.55$, and $\Lambda=0.183333$. The radii are $a_{T1}=1.42774$ and $a_{T2}=9.57226$.} \end{figure}

 Since   $a_{1}>0$ and $a_{2}>0$ require $\Delta=\frac{1}{3}(3+c_4(6-4c_3)-c_4^2)>0$, $c_3 + c_4<0$, and $-(c_4 + 2 c_3-1)>0$,  the conditions for two positive roots are
\begin{eqnarray}\label{cubic02+1}
&& c_3<1 \;,\;\;3-c_3-2\sqrt{3-3c_3+c_3^2}<c_3+c_4<0\;,
\nonumber\\
&& 2<c_3\leq 3 \;,\;\;3-c_3-2\sqrt{3-3c_3+c_3^2}<c_3+c_4<3-2c_3\;,
\nonumber\\
or && c_3>3 \;,\;\;3-c_3-2\sqrt{3-3c_3+c_3^2}<c_3+c_4<6-3c_3\;.
\end{eqnarray}
In Region 1 of Fig.~(\ref{Figurecubic0-}), we show the allowed values $c_3$ and $c_4$ for two positive roots.
 Setting $a_1=a_{T2}$ and $a_2=a_{T1}$ since $a_1>a_2$, we show the evolution of the potential ${V}(a)$  in Fig.~(\ref{Figurez3})  with the model parameters in Region 1 of
Fig.~(\ref{Figurecubic0-}). From this figure, we see that a $BB\Rightarrow BC$
universe or a bouncing one is
obtained since $V\leq 0$ in $a\in (0, a_{T1}]$ and $[a_{T2},\infty)$ with $V=0$ occurring   at $a=a_{T1}$ and $a=a_{T2}$. Another possibility is that the universe starts at the big bang singularity and, rather than bounce back, it quantum mechanically tunnels to $a_{T1}$ when it evolves to $a_{T2}$ and then expands forever.

\begin{figure}[htbp]
\includegraphics[width=7cm]{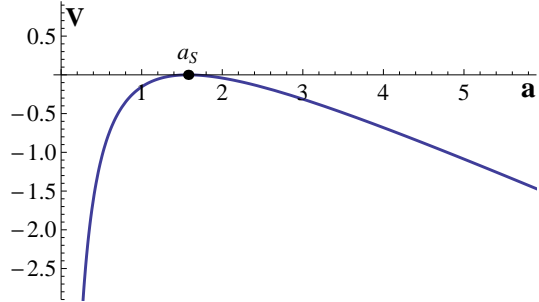}
\caption{\label{Figurez4}  Potential $V(a)$ with model parameters satisfying Eq.~(\ref{cubic02+1}). An unstable Einstein static state universe is obtained. The constants are chosen as $m=1$, $c_3=3$, $c_4=-4.64575$, and $\Lambda=0.15470$. $a_s=1.57735$.}
\end{figure}

If $c_4=3-2c_3\pm2\sqrt{3+c_3(c_3-3)}$,  it is easy to see $a_{1}=a_{2}$, which means that $V(a)=0$ has a double solution $a_s$
\begin{eqnarray}\label{cubic0double-}
a_{s}=\frac{-3+2c_3+2\sqrt{3+c_3(c_3-3)}}{3(c_3-2)}\;.
\end{eqnarray}
  We  find that $a_s$ is an unstable ES universe and it exists under the conditions
\begin{eqnarray}\label{cubic02+1}
 && c_3<1\;,\;\;c_3+c_4=3-c_3-2\sqrt{3+c_3(c_3-3)}\;,\nonumber \\
 or  && c_3>2\;,\;\;c_3+c_4=3-c_3-2\sqrt{3+c_3(c_3-3)}\;.
\end{eqnarray}
Thus, as shown in Fig~(\ref{Figurez4}), if the universe initiates from big bang, it will expand to an unstable ES universe and then turn over or  expand forever.

\begin{figure}[htbp]
\includegraphics[width=7cm]{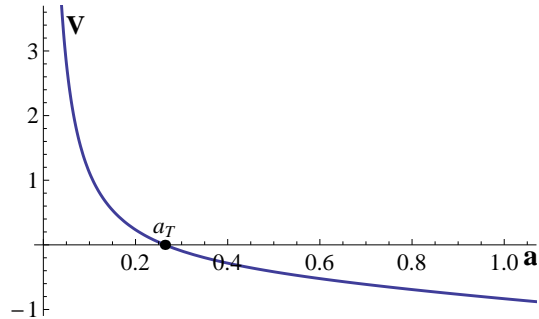}
\caption{\label{Figurez5}   Potential $V(a)$  with model parameters in Region 1 of Fig.~(\ref{Figurecubic0-}). A bouncing universe is found.  The constants are chosen as $m=1$, $c_3=1$, $c_4=-0.5$, and $\Lambda=0.83333$. The turning radius at a bounce is $a_T=0.263763$.}
\end{figure}

Now we consider the case of  $a_{2}<0$ and $a_{1}=a_T>0$. In this case,  a bouncing universe is obtained, as shown in Fig.~(\ref{Figurez5}),  if  $c_3$ and $c_4$ satisfy
\begin{eqnarray}\label{cubic02+1}
c_3<\frac{3}{2} \;,\;\;0<c_3+c_4<3-2c_3,
\end{eqnarray}
which corresponds to Region $2$ of Fig.~(\ref{Figurecubic0-}).

\begin{figure}[htbp]
\includegraphics[width=7cm]{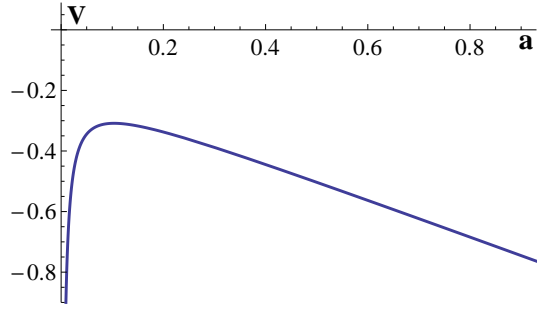}\quad
\caption{\label{Figurez6} Potential ${V}(a)$ for the case of without positive root.
 A $BB\Rightarrow\infty$ or $\infty\Rightarrow BC$ one is obtained.  The constants are
chosen as $m=1$, $c_3=1.2$, $c_4=-1.22$, and $\Lambda=0.806667$.}
\end{figure}

If  both $a_{1}$ and $a_{2}$ are not positive,  we find that the potential is always negative as shown in Fig.~(\ref{Figurez6}) and the cosmic type is $BB\Rightarrow\infty$ or $\infty\Rightarrow BC$.

 The results of this section are summed up in Tab.~(\ref{Tab3}).

\begin{table}[!h]
\tabcolsep 3pt \caption{\label{Tab3} Summary of the cosmic type in the  $4c_3+c_4-6+3\Lambda=0$ case} \vspace*{-12pt}
\begin{center}
\begin{tabular}{|c|c|c|c|c|c|}
  \hline
  $c_3, \; c_4$                    & Cosmic Type\\ \hline
  {$  c_3\leq \frac{3}{2},\;\;3-2c_3<c_3+c_4<6-3c_3 $}      &  Oscillation  \\
  {$or~ \frac{3}{2}<c_3<2,\;\;0<c_3+c_4<6-3c_3 $}      &    \\ \hline
  {$\frac{3}{2}<c_3\leq 2,\;\;3-2c_3<c_3+c_4<0$}  &  $BB\Rightarrow BC$\\
  {$or~2<c_3<3,\;\;3-2c_3<c_3+c_4<6-3c_3$}  &     \\ \hline
 {$  c_3<1,\;\;3-c_3-2\sqrt{3-3c_3+c_3^2}<c_3+c_4<0 $}      &  $BB\Rightarrow BC$  \\
   {$2<c_3\leq 3,\;\;3-c_3-2\sqrt{3-3c_3+c_3^2}<c_3+c_4<3-2c_3 $}       & or Bounce \\
  {$or~  c_3>3,\;\;3-c_3-2\sqrt{3-3c_3+c_3^2}<c_3+c_4<6-3c_3$}  &         \\ \hline
                                                    &              $BB\Rightarrow BC$\\
    {$c_3<1,\;\;c_3+c_4=3-c_3-2\sqrt{3+c_3(c_3-3)}$}     &         $BB\Rightarrow \infty$  \\
   {$or~c_3>2,\;\;c_3+c_4=3-c_3-2\sqrt{3+c_3(c_3-3)}$}                 &$\infty\Rightarrow BC$ \\
                                                                 & or Bounce\\
                                                           & Unstable ES\\ \hline
  {$c_3<\frac{3}{2},\;\;0<c_3+c_4<3-2c_3,$}  &              Bounce\\

  \hline
\end{tabular}
     \end{center}
       \end{table}

\section{Conclusions}
Massive gravity is a modification of general relativity. It has been spurring  an increasing deal of interest recently, since it can explain the present accelerated cosmic expansion without the need of dark energy.  In this paper, using a method in which the scale factor $a$ changes as a particle in a ``potential", we analyze  all possible cosmic evolutions in a ghost-free massive gravity theory.   A spatially flat universe is considered in our discussion and we assume that the vacuum energy is the only energy component. The results are summed up in Tabs.~(\ref{Tab1}, \ref{Tab2}, \ref{Tab3}). We find that
 there may exist an oscillating universe between $a_{min}$ and $a_{max\;}$
or a bouncing one at $a_T$ if model
parameters are in some specific regions.
If the cosmic scale factor is in the
region $[a_{min}, a_{max\;}]$ initially, the universe may undergo an
oscillation. After a number of oscillations, it may evolve to the
bounce point $a_T$ through quantum tunneling. While, if the
universe contracts initially from an infinite scale, it can turn
around at $a_T$ and then expand forever.   Thus, the big bang
singularity problem can be  resolved successfully.
  Remarkably, although we do have a stable ES solution
 in some circumstances, the universe can not stay at this stable state past-eternally  since it is allowed to
 quantum mechanically tunnel to a big-bang-to-big-crunch  cosmic evolution type and end with a big crunch. Thus, the existence of a stable ES universe  can not successfully resolve the big bang singularity in the massive gravity.  This feature is related to the behavior of  $V(a)\rightarrow-\infty$ as $a\rightarrow 0$, which is a result of $c_3+c_4< 0$ in the massive gravity we consider in the present paper, when there exists a stable ES universe  with a finite $a_s$.  Let us note however that both in the Horava-Lifshitz gravity~\cite{Maeda2010}  and  the DGP braneworld scenario~\cite{Zhang2012}, there exist stable ES universes where $V(a)\rightarrow\infty$ as $a\rightarrow 0$, so a quantum tunneling to the big bang singularity is not allowed ( refer to Fig.~(4) in \cite{Maeda2010}  and Fig.~(5) in \cite{Zhang2012})  and as a result  the existence of a stable ES universe  can avoid  the big bang singularity in these theories. Therefore, whether the existence of a stable ES universe can resolve the big bang singularity or not is a peculiar feature of the  theory of gravity itself and it is in fact determined by the behavior of the leading term in $V(a)$ as $a$ approaches zero, i.e., the big bang singularity.

\acknowledgments
This work was supported by the National Natural Science Foundation of China under Grants Nos. 10935013, 11175093, 11222545 and 11075083, Zhejiang Provincial Natural Science Foundation of China under Grants Nos. Z6100077 and R6110518, the FANEDD under Grant No. 200922, the National Basic Research Program of China under Grant No. 2010CB832803, the NCET under Grant No.  09-0144,  the PCSIRT under Grant No. IRT0964, the SRFDP under Grant No. 20124306110001, the Hunan Provincial Natural Science Foundation of China under Grant No. 11JJ7001,  the SRFDP under Grant No. 20124306110001,  the Program for the Key Discipline in Hunan Province, and
Hunan Provincial Innovation Foundation For
Postgraduate under Grant No. CX2012B203.


\end{document}